\documentclass[sigconf]{acmart}

\usepackage{pifont} 
\newcommand{\xmark}{\ding{55}}
\usepackage{makecell}  % 放在导言区
\usepackage{hyperref}       % hyperlinks
\usepackage{url}            % simple URL typesetting

\AtBeginDocument{%
  }

\setcopyright{acmlicensed}
\copyrightyear{2018}
\acmYear{2018}
\acmDOI{XXXXXXX.XXXXXXX}
\acmConference[Conference acronym 'XX]{Make sure to enter the correct
  conference title from your rights confirmation email}{June 03--05,
  2018}{Woodstock, NY}
\acmISBN{978-1-4503-XXXX-X/2018/06}

\begin{document}

\title{DS4RS: Community-Driven and Explainable Dataset Search Engine for Recommender System Research}

\author{Xinyang Shao}
\email{xinyang.shao@huawei-partners.com}
\affiliation{%
  \institution{Huawei Ireland Research Centre}
  \city{Dublin}
  \country{Ireland}
}

\author{Tri Kurniawan Wijaya}
\email{tri.kurniawan.wijaya@huawei.com}
\affiliation{%
  \institution{Huawei Ireland Research Centre}
  \city{Dublin}
  \country{Ireland}
}

\begin{abstract}
Accessing suitable datasets is critical for research and development in recommender systems. However, finding datasets that match specific recommendation task or domains remains a challenge due to scattered sources and inconsistent metadata. To address this gap, we propose a community-driven and explainable dataset search engine tailored for recommender system research. Our system supports semantic search across multiple dataset attributes, such as dataset names, descriptions, and recommendation domain, and provides explanations of search relevance to enhance transparency. The system encourages community participation by allowing users to contribute standardized dataset metadata in public repository. By improving dataset discoverability and search interpretability, the system facilitates more efficient research reproduction. The platform is publicly available at: \url{https://ds4rs.com}.

\end{abstract}




\maketitle

\section{Introduction}

Recommender systems play a central role in modern digital services, supporting personalized experiences in domains such as e-commerce, streaming media, and social platforms~\cite{Roy2022,10.1109/TKDE.2005.99}. As the field continues to evolve, the ability to access suitable datasets becomes increasingly important for training, benchmarking, and reproducing new models~\cite{Beel2016, 10.1145/2827872}. However, researchers often struggle to locate datasets that align with their specific task objectives—such as click-through rate (CTR) prediction, rating prediction, or Top-N recommendation—due to fragmented sources and the lack of standardized metadata~\cite{Chapman2020,9843966}.

Although online search engine (such as Google and Bing), 
public repositories~\cite{mcauley_datasets,xwines2023,evalrs2023}, and 
static websites~\cite{rucaibox_recsysdatasets} offer broad access to dataset-related content, 
and platforms like Kaggle\cite{kaggle} 
and Hugging Face Datasets~\cite{huggingface} 
provide downloadable datasets, 
these solutions are mostly general-purpose in nature and 
might not be tailored to the specific needs of recommender system research~\cite{Akhtar_2024,9843966}. 
Additionally, they may lack structured classification of recommendation tasks and 
provide limited metadata about dataset schema, 
example content, or recommendation domain~\cite{10.1145/3477495.3536334}. 
As a result, it might be difficult for researchers to efficiently discover datasets aligned with their modeling objectives or to interpret their relevance for specific applications.

To address these challenges, we introduce a \textit{community-driven} and \textit{explainable} dataset search engine specifically designed for the recommender systems domain\footnote{A demonstration video is available at \url{https://bit.ly/ds4rs-video}.}. The search engine supports semantic retrieval across multiple metadata fields, including dataset names, general descriptions, recommendation domian and so on. Furthermore, an interactive interface has been developed to support dataset exploration, metadata inspection, and direct access to download links. This system aims to improve dataset discoverability~\cite{10.1007/s00799-022-00339-w}, enhance transparency in dataset selection, and accelerate research reproducibility~\cite{10.1007/s11257-021-09302-x} in the recommender systems community.

\section{System Architecture}

\begin{figure}[htbp]
  \centering
  \includegraphics[width=1\linewidth]{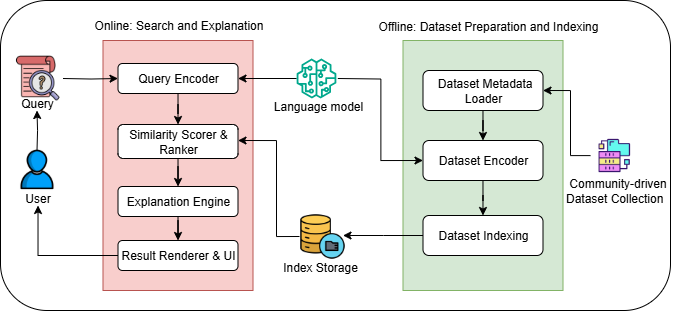}
  \caption{DS4RS System Architecture}
  \label{fig:engine-arch}
\end{figure}

Our system consists of three main components: a community-driven dataset repository, an offline indexing pipeline, and an online semantic search and explanation module. Together, they form an integrated and explainable search engine tailored for recommender system datasets~\cite{Chapman2020}, as illustrated in Figure~\ref{fig:engine-arch}. In the following, we describe two key features that enable community extensibility and explainable retrieval.

\subsection{Community-Driven Metadata Contribution}

A core component of our architecture is the community-driven dataset repository, which enables flexible and scalable metadata collection. Each dataset is described using a structured JSON schema that captures recommendation task (CTR prediction, rating prediction, Top-N recommendation), application domain (e.g., movie, e-commerce), dataset size, textual description, record example. These JSON files are maintained in our public repository~\cite{ds4rs-repo}, 
allowing researchers to contribute new datasets in a standardized format. 
This community-driven approach not only enables scalable and decentralized dataset collection, 
but also increases the awareness and reusability of contributed datasets.

\subsection{Explainable Semantic Search}

During the offline pre-processing stage, all submitted metadata files are loaded, validated, and encoded in advance. The metadata fields of the dataset are independently encoded using the Sentence-BERT model, \texttt{all-mpnet-base-v2}, which is fine-tuned for semantic similarity tasks and has achieved state-of-the-art results in sentence-level semantic representation~\cite{reimers2019sentencebertsentenceembeddingsusing,sentence-transformers-pretrained-models}. This design not only ensures consistency across metadata fields, but also improves online search efficiency by decoupling costly encoding operations from real-time query handling. 

The retrieval module performs Semantic search over multiple metadata fields, including dataset name, description, recommendation task and domain. To support transparency, the user query is encoded in the same vector space and compared with all dataset fields using cosine similarity. The final relevance score is defined as the maximum similarity across fields. The system provides field-level explanations by showing which metadata fields contributed most to each match, along with their similarity scores. This allows users to better interpret the ranking results and make more informed dataset selections.

\section{Use Cases}

To demonstrate the usability and effectiveness of our system, we developed an interactive prototype that supports semantic search, relevance explanation, and field-level dataset exploration. 

\begin{figure}[htbp]
  \centering
  \includegraphics[width=\linewidth]{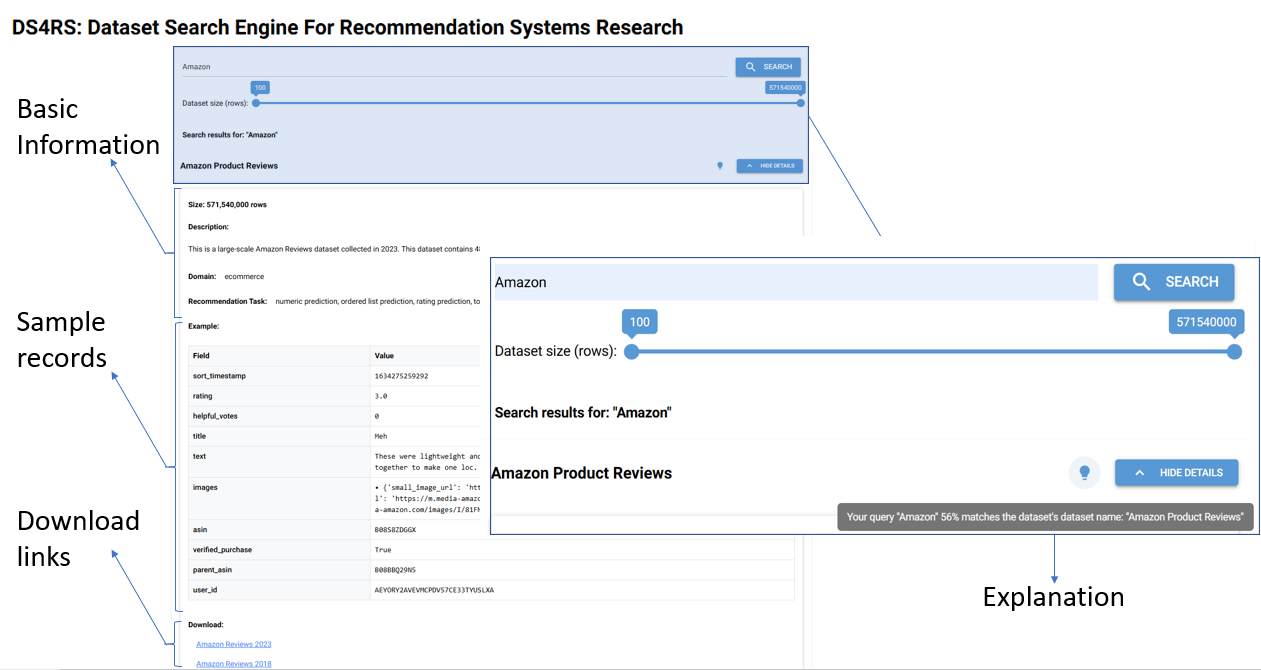}
  \caption{Dataset search interface showing ranked and expandable results}
  \label{fig:ui-example}
\end{figure}

% \vspace{0.2em}
% \noindent
% \textbf{Explainable Dataset search engine} 
\textit{Explainable dataset search engine.}
Users can then interact with the system through a clean web interface, where they enter free-form queries to retrieve relevant datasets. The system returns a ranked list of datasets annotated with recommendation tasks and associated domains. The user can further apply a size filter to narrow down results and expand result cards to view the full metadata, including example records and download links. This interface is illustrated in Figure~\ref{fig:ui-example}. 
To enhance transparency, each explanation identifies the most influential fields, enabling users to understand the retrieval rationale and assess dataset relevance to their tasks.

\textit{Community-driven public repository.}
Researchers can contribute new datasets by uploading metadata files to our public repository~\cite{ds4rs-repo}
using a standardized JSON schema we designed. 
Once validated and uploaded, the contributed metadata becomes part of the indexed collection, enabling its discovery through subsequent semantic queries.

To contextualize the functionality, Table~\ref{tab:comparison} compares the support of key features between the existing solutions and ours.

\begin{table}[h]
  \centering
  \caption{Feature comparison across various existing solutions and ours (DS4RS).}
  \label{tab:comparison}
  \resizebox{\linewidth}{!}{
  \begin{tabular}{p{4.3cm}cccc}
    \hline
    \textbf{Features} & \makecell[c]{Generic, Online\\ Search Engine} & \makecell[c]{Public Repositories \\ and Static Websites} & \makecell[c]{Kaggle / \\ Hugging Face} & DS4RS \\
    \hline\hline
    Syntactic search                    & \checkmark     & \checkmark & \checkmark & \checkmark \\
    Semantic search                     & \checkmark     & \xmark     & \xmark     & \checkmark \\
    Dataset link                        & \checkmark     & \checkmark & \checkmark & \checkmark \\
    Retrieval by recommendation task \newline (e.g., CTR prediction) & \checkmark     & \xmark     & \xmark     & \checkmark \\
    Filter by dataset size              & \xmark         & \xmark     & \checkmark & \checkmark \\
    Result relevance explanation        & \xmark         & \xmark     & \xmark     & \checkmark \\
    Open source dataset collection      & \xmark         & \checkmark & \checkmark     & \checkmark \\
    \hline\hline
  \end{tabular}
  }
\end{table}

This comparison underscores a key distinction: while existing tools support general dataset access, they are not specialized for recommender systems. As a result, queries on more generic platforms (e.g., Google, Kaggle) 
may yield loosely related datasets lacking user-item interactions or task annotations. DS4RS, by contrast, indexes only datasets explicitly associated with recommendation tasks, ensuring higher relevance and direct applicability to typical modeling objectives such as CTR prediction or top-N ranking. This design focus improves the efficiency and reliability of dataset discovery for recommendation research.

\section{Conclusions and Future Work}

We presented a community-driven and explainable dataset search engine tailored for recommender system research by combining structured metadata, semantic field-aware retrieval, and an interactive user interface to make dataset discovery, research reproducibility, and methodological benchmarking easier within the recommendation community.

For future work, we plan to introduce a dataset comparison feature, allowing users to compare datasets side-by-side in terms of recommendation tasks, field coverage, and data scale. 
We also aim to provide more customizable filters, enabling users to perform more accurate and fine-grained searches based on their specific needs. 
Lastly, we will continue to enhance our community contribution mechanisms, including contributor gamification and seamless user feedback integration, to foster an open and collaborative ecosystem. 
We invite researchers and practitioners to contribute new datasets and suggestions through our public repository~\cite{ds4rs-repo}
and help shape a shared resource for the recommendation community. 

\bibliographystyle{ACM-Reference-Format}
\bibliography{reference}
 
\end{document}